\documentclass[5p,twocolumn]{elsarticle}

\usepackage{amssymb}
\usepackage{amsmath}
\usepackage{graphicx}
\usepackage{verbatim}
\usepackage[colorlinks,hyperindex]{hyperref}

\usepackage[utf8x]{inputenc}
\usepackage{xparse}
\usepackage{physics}
\usepackage{kantlipsum}
\usepackage{color}

\usepackage{float}

\journal{Physics Letters A}

\begin{document}

\begin{frontmatter}

\title{Indirect strong coupling regime between a quantum dot and a nanocavity mediated by a mechanical resonator}
\author[]{J. E. Ramírez-Muñoz\corref{cor1}}
\ead{jeramirezm@unal.edu.co }
\author[]{J. P. Restrepo Cuartas}
\author[]{H. Vinck-Posada}
\address[]{Universidad Nacional de Colombia - Sede Bogotá, Facultad de Ciencias, Departamento de Física\\
 Carrera 45 No. 26-85, C.P. 111321, Bogotá, Colombia}
\cortext[cor1]{Corresponding authors}
\date{\today}

\begin{abstract}
Achieving strong coupling between light and matter is usually a challenge in Cavity Quantum Electrodynamics (cQED), especially in solid state systems. For this reason is useful taking advantage of alternative approaches to reach this regime, and then, generate reliable quantum polaritons. In this work we study a system composed of a quantized single mode of a mechanical resonator interacting linearly with both a single mode nanocavity and a two-level single quantum dot. In particular, we focus on the behavior of the indirect light-matter interaction when the phonon mode interfaces both parts. By diagonalization of the Hamiltonian and computing the density matrix in a master equation approach, we evidence several features of strong coupling between photons in optical cavities and excitons in a quantum dot. For large energy detuning between the cavity and the mechanical resonator it is obtained a phonon-dispersive effective Hamiltonian which is able to retrieve much of the physics of the conventional Jaynes-Cummings model (JCM). In order to characterize this mediated coupling, we make a quantitative comparison between both models and analyze light-matter entanglement and purity of the system leading to similar results in cQED.
\end{abstract}


\end{frontmatter}


\section{Introduction}

In the last decades solid state systems have been widely used as a platform to interface light with matter in the cavity quantum electrodynamics (cQED) frame \cite{Lodahl2015,Khitrova2006,Laussy2007a} which opens the route towards quantum computation and quantum information processing devices \cite{Nielsen2000}. Different regimes of light-matter interaction have been studied and each of them are useful to play some role in current technologies. On the one hand, weak coupling between light and matter can be used to generate single-photons on-demand; on the other hand, strong coupling regime is exploited to transfer quantum information between light and matter and perform quantum logic operations. Additionally, the control over the light-matter interaction rate and the spontaneous emission decay are a desired feature in order to achieve versatility and efficiency. 

Since quantum dot-microcavity strong coupling  is difficult to achieve in many cQED systems due to fabrication issues and incoherent processes \cite{Reithmaier2004,Yoshie2004}, it is really useful taking advantage of other mechanisms that allow to reach this regime. Mechanical modes of semiconductor structures and surface acoustic waves (SAW) are now used to handle the properties of quantum dots embedded in optical microcavities \cite{Lima2005,Pennec2014}. Usually, the interaction of quantized mechanical modes \cite{Schwab2005} with quantum dots and optical cavities are modeled with electron-phonon Hamiltonian \cite{Wilson-Rae2004,Yeo2014} and optomechanical Hamiltonian \cite{Law1995,Aspelmeyer2014}, respectively. Large optomechanical coupling rates and photon-phonon strong coupling regime have been achieved in different systems, particularly in photonic crystals and micropillars \cite{Gavartin2011,Fainstein2013}. Besides, active media have been used to enhance optomechanical coupling \cite{Pirkkalainen2015,Jusserand2015,Rozas2014}. 

Furthermore, it is well known that linear coupling between cavity and mechanical resonators raises in the so-called \textit{resolved sideband regime} \cite{Verhagen2012,Aspelmeyer2014,Groblacher2009}, i.e., when phonon energy is larger than the cavity decay rate ($\omega_{m}>\kappa$), in that regime the cavity is pumped by a large coherent field. In addition, there has been recently proposed a coherent field-mediated emitter-phonon linear coupling produced by variations of the spatial distributions of the cavity field \cite{Cotrufo2017,Xiang2013}. The aim of this work is to study alternative models in polariton cavity optomechanics, different from the standard hybrid optomechanical Hamiltonians in the dispersive regime \cite{Restrepo2017,Kyriienko2014,Restrepo2014,Wang2015,Zhou2016}. This alternative way is exploited to obtain enhancement of light-matter coupling rate and its quantum features such as entanglement which is a key resource for quantum information operations \cite{Horodecki2009,Blattmann2014,Vera2009,Suarez-Forero2016}. However, other scenarios involving acoustic phonons show a detrimental impact on the light-matter strong coupling and its effects \cite{Glassl2012,Holz2015}.

The rest of the paper is organized as follows, in section \ref{sec:theory} we set the theoretical model for the tripartite cavity-emitter-mechanics system. Then, in section \ref{sec:strong_coupling}, by diagonalization of the Hamiltonian we discuss the possibility of achieving strong coupling regime mediated by phonons and study the dispersive limit of the Hamiltonian. In section \ref{sec:entanglement}, using the density operator in the steady state we quantify light-matter entanglement and linear entropy, and show the tomography in different important situations. Finally, in section \ref{sec:conclusions} we make and overview and conclude.

\section{Theoretical framework}
\label{sec:theory}

\begin{figure}[H]
\centering
{\includegraphics[scale=0.6]{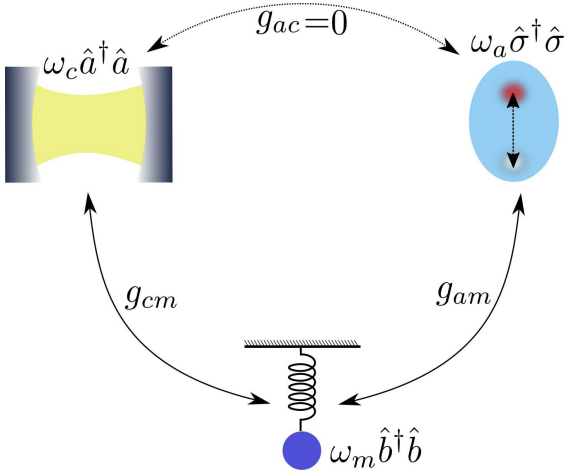}}
\caption{Sketch of the hybrid cQED-optomechanical system \cite{Restrepo2017}. Linear coupling is assumed between the nanocavity and the mechanical resonator, and also between the single-exciton quantum dot and  the mechanical resonator. Throughout this work there is no direct interaction between the nanocavity and the quantum dot.}
\label{system}
\end{figure}

As stated above, the tripartite system in Figure \ref{system} is typically modeled with optomechanical and electron-phonon dispersive interactions. However, we will alternatively use an approximate Hamiltonian; linear coupling between different parts of the system as is justified in the appendix \ref{app:linearization}. Here, we do not consider light-matter interactions directly. Thus, the Hamiltonian is:

\begin{equation}
\hat{H}=\hat{H}_{0}+\hat{H}_{int}
\label{total}
\end{equation}

\noindent with

\begin{equation}
\hat{H}_{0}=\hbar\omega_{c}\hat{a}^{\dagger}\hat{a}+\hbar\omega_{a}\hat{\sigma}^{\dagger}\hat{\sigma}+\hbar\omega_{m}\hat{b}^{\dagger}\hat{b},
\end{equation}

\noindent and

\begin{equation}
\hat{H}_{int}=\hbar g_{cm} \left(\hat{a}\hat{b}^{\dagger}+\hat{a}^{\dagger}\hat{b} \right)+\hbar g_{am} \left(\hat{\sigma}\hat{b}^{\dagger}+\hat{\sigma}^{\dagger}\hat{b} \right).
\label{linear_coupling}
\end{equation}

\noindent $\omega_c$, $\omega_a$ and $\omega_m$ are the cavity, exciton recombination and mechanical frequencies, respectively. $\hat{a}$ ($\hat{a}^{\dagger}$) and $\hat{b}$ ($\hat{b}^{\dagger}$) are the annihilation (creation) bosonic operators for photons and phonons, respectively. $\hat{\sigma}=\ket{G}\bra{X}$ is the atomic ladder operator for the two-level system (TLS). The parameters $g_{cm}$ and $g_{am}$ are the optomechanical coupling strength and the atom-phonon coupling strength, respectively.

Losses and decoherence arise in the system when it is affected by its environment. These are effectively considered in a master equation approach for the density operator in the Markov approximation \cite{Walls2007,Scully1997}:

\begin{eqnarray}
\frac{d\hat{\rho}}{dt}=\frac{i}{\hbar}\left[\hat{\rho}, \hat{H} \right]+\kappa \mathcal{L}_{\hat{a}}(\hat{\rho})+P_{X}\mathcal{L}_{\hat{a}^{\dagger}}(\hat{\rho}),
\label{master_equation}
\end{eqnarray}

\noindent where $\kappa$ is the cavity decay rate and $P_{x}$ is an incoherent excitation to the quantum dots. The Lindblad term for the operator representing a dissipative channel, $\hat{C}$, is defined as $\mathcal{L}_{\hat{C}}(\hat{\rho})=\hat{C}\hat{\rho}\hat{C}^{\dagger}-\frac{1}{2}\hat{C}^{\dagger}\hat{C}\hat{\rho}-\frac{1}{2}\hat{\rho}\hat{C}^{\dagger}\hat{C}$.

Throughout this work, we assume high mechanical coupling rates, $g_{am}/\omega_a\approx g_{cm}/\omega_c  > 10^{-3}$, which could be obtained taking advantage of large coherent driven as is shown in the appendix \ref{app:linearization}. While mechanical frequencies considered here are very out of resonance with cavity frequencies ($\omega_m/\omega_c\sim 10^{-3}$), cavity and exciton recombination frequencies are close each other, i.e. $|\Delta|=|\omega_{a}-\omega_{c}|<g_{am}$. Assuming realistic parameters of the Microcavity Quantum Electrodynamics ($\mu$cQED), $\omega_{c}\approx\omega_{a}\approx 1000meV$, such that $g_{cm}\approx g_{am}\sim meV$. From here on, all frequencies are given in $meV$, assuming $\hbar=1$.

\begin{figure}[H]
\centering
{\includegraphics[scale=0.4]{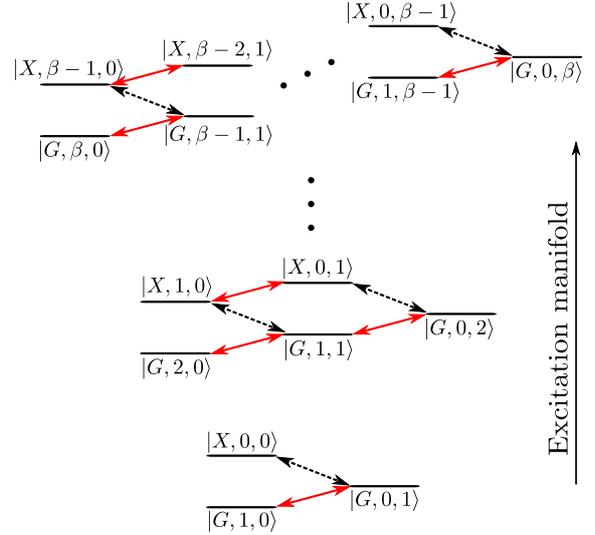}}
\caption{States ladder of the Hamiltonian. Black-dashed arrows represent exciton-phonon interactions ($g_{am}$) and the red ones correspond to photon-phonon interactions ($g_{cm}$). A $\beta$-excitation manifold has $2\beta+1$ states $\lbrace \left|G,\beta-k,k \right>, \left|X,\beta-k-1,k \right>\rbrace$, $k=0,1,...,\beta-1$.}
\label{ladder}
\end{figure}

\section{Indirect light-matter strong coupling regime}
\label{sec:strong_coupling}

Strong coupling between quantum dots and optical cavities is a hard feature to achieve; usually it requires high control in the positioning of QD inside the cavity, small field modal volume and high quality factor. However, indirect mechanics can be harnessed to reach the vacuum Rabi splitting regime. We consider a quantized mechanical mode interacting simultaneously with both the cavity and QD in a linear way according to equation (\ref{linear_coupling}). Since the Hamiltonian commutes with the total number operator ($N=\hat{a}^{\dagger}\hat{a}+\hat{\sigma}^{\dagger}\hat{\sigma}+\hat{b}^{\dagger}\hat{b}$), then it can be diagonalized for each excitation manifold as shown in Figure \ref{ladder}.

\begin{figure}[H]
\centering
\includegraphics[scale=0.14]{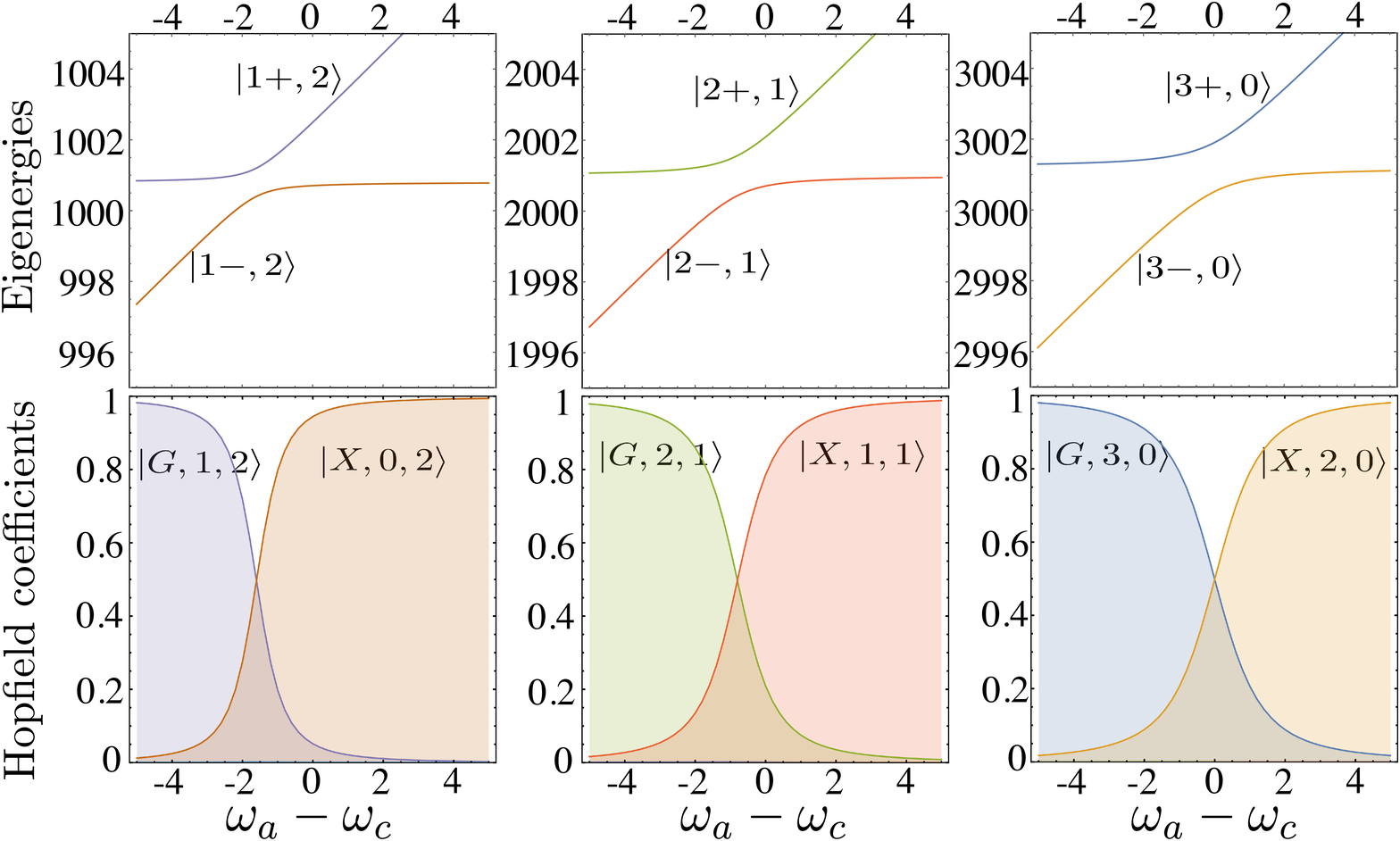}
\caption{Numerical diagonalization of the Hamiltonian for the third- excitation manifold. The trivial phonon eigenstate is the bare state $\left|G,0,3\right>$ with eigenenergy $\omega_m$ (not shown here). Parameters: $g_{cm}=g_{am}=20$, $\omega_m=1$ and $\omega_c=1000$.}
\label{third-manifold}
\end{figure}

By numerical diagonalization of the Hamiltonian in a large energy detuning regime between cavity (emitter) and mechanical oscillator, is found that the eigenstates of the system are separable in a polariton state and a phonon Fock state: $\left|\psi_{n}^{\ell}\right>\approx\left|n\pm\right>\otimes\left|\ell\right>$. From here on, polariton in this context will be understood as a \textit{phonon-induced polariton} (PIP). Moreover, similarly to the Jaynes-Cummings model, the dispersion diagram exhibits anticrossing between two dressed states; the upper and lower polaritons, $\left|n+\right>\otimes\left|\ell\right>$ and $\left|n-\right>\otimes\left|\ell\right>$. Each polariton is a superposition of light-matter bare states: $\left|n\pm\right>=\cos{\theta_{n}}\left|G,n\right>\pm \sin{\theta_{n}}\left|X,n-1\right>$, where $\theta_n$ depends on the system parameters and photon number. This situation is illustrated in the Figure \ref{third-manifold} for the third-excitation manifold. 

In the limit of large cavity-photon detuning \cite{Gerry2005c}, ($|\Delta_c|=|\omega_c-\omega_m|\gg 1$), and emitter-cavity resonance ($\omega_{a}=\omega_{c}$), as is derived in the appendix \ref{app:dispersive}, it can be obtained an effective Jaynes-Cummings Hamiltonian:

\begin{eqnarray}
\hat{H}_{eff}=\frac{g_{cm}^{2}}{\Delta_{c}}\hat{a}^{\dagger}\hat{a} &+& \frac{g_{am}^{2}}{\Delta_{c}}\hat{\sigma}^{\dagger}\hat{\sigma}+\left(\frac{g_{am}^{2}}{\Delta_{c}}\sigma_z - \frac{g_{cm}^{2}}{\Delta_{c}} \right) \hat{b}^{\dagger}\hat{b} \nonumber\\
& &\frac{g_{cm}g_{am}}{\Delta_c}\left( \hat{a}\hat{\sigma}^{\dagger} + \hat{a}^{\dagger}\hat{\sigma} \right).
\label{effective_hamiltonian}
\end{eqnarray}

\noindent This Hamiltonian is block-diagonal respect to the phonon number and for each one, $\left<\ell\right|H_{eff}\left|\ell\right>$ is a block-diagonal matrix respect to the polaritonic excitation manifold, $\lbrace\left|G,n\right>,\left|X,n-1\right>\rbrace$, in a similar way that the standard Jaynes-Cummings Hamiltonian:

\begin{eqnarray}
\hat{H}_{eff}=
\begin{pmatrix} 
\frac{g_{cm}^{2}}{\Delta_c}n-\frac{g_{am}^{2}}{\Delta_c}\ell & \frac{g_{cm}g_{am}}{\Delta_c}\sqrt{n} \\
\frac{g_{cm}g_{am}}{\Delta_c}\sqrt{n} & \frac{g_{cm}^{2}}{\Delta_c}(n-1)-\frac{g_{am}^{2}}{\Delta_c}(\ell+1)
\end{pmatrix}
\end{eqnarray}

\noindent By diagonalization of the previous matrix, it is obtained some useful relations. For each pair of upper/lower phonon-induced polaritons the anticrossing arises in $\omega_a-\omega_c=\bar{g}_{m}\ell$ where

\begin{equation}
\bar{g}_{m}=\frac{g_{am}^{2}+g_{cm}^{2}}{\omega_a-\omega_m}
\end{equation}

\noindent is the effective light-matter coupling strength. The splitting at resonance is 

\begin{equation}
\lambda_{+}-\lambda_{-}=\frac{g_{cm}^{2}}{\Delta_{c}}\sqrt{1+\epsilon^{4}(1+2\ell)^{2}+2\epsilon(2n-2\ell-1)},
\end{equation}

\noindent where $\epsilon=g_{am}/g_{cm}$ is the ratio between the mechanical interaction rates. Moreover, in the anticrossing region, $\ell=0$ and $\epsilon=1$, the splitting between upper and lower polaritons is $\bar{g}_{m}\sqrt{n}$ in analogy with the Jaynes-Cummings case. 

\begin{figure}[H]
\centering
\includegraphics[height=8cm, width=7.5cm]{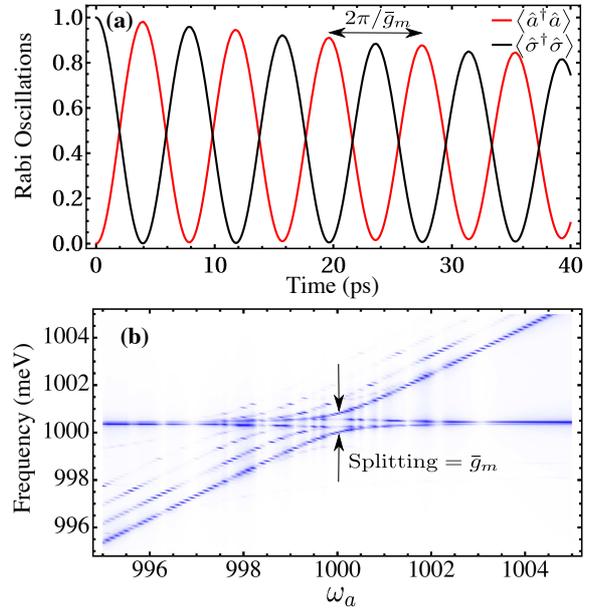}
\caption{Vacuum Rabi Splitting. (a) Mean number of photons (Red line) and matter excitations (Black line). They display damped Rabi oscillations with frequency $\bar{g}_{m}$. (b) Cavity light-emitted spectrum computed by means of the Green function method \cite{Gomez2015}. The spectrum is normalized to unity in order to show other optical transitions. A splitting of $\bar{g}_{m}$ is obtained. Parameters: $g_{cm}=g_{am}=20$, $\omega_{c}=1000$, $\omega_{m}=1$, $\kappa=10^{-2}$ and $P_{X}=10^{-3}$. Time units $meV^{-1}/\hbar\approx 1.5^{-1}ps$.}
\label{spectrum}
\end{figure}

Summarizing, the Hamiltonian proposed (Equation (\ref{total})) allows a major control on the effective light-matter coupling strength, e.g., by increasing the mechanical interaction rates or decreasing the cavity (exciton) energy, the splitting could become larger than the spectral linewidths, i.e., $\bar{g}_m>\kappa$. Hence, the emission exhibits Vacuum Rabi splitting showing a quantum coherent exchange of particles between the nanocavity and the quantum dot, which gives a first feature of strong coupling regime. We can observe this behavior in the mean number of photons and matter excitations since they display Rabi oscillations (Figure \ref{spectrum}a), and also in the photoluminescence spectrum which is defined as follows:

\begin{equation}
S_{a}(\omega,t)=\frac{1}{\pi \left<a^{\dagger}a \right>}Re \int_{0}^{\infty}e^{-i\omega \tau}\left<a^{\dagger}(t+\tau)a(t)\right>d\tau.
\label{stimulated_spectrum}
\end{equation}

\noindent Emission cavity spectrum shown in Figure \ref{spectrum}b exhibits single-polariton optical transitions for each phonon number, i.e., $\ket{1\pm,\ell}\rightarrow\ket{G,0,\ell}$. Optical transitions involving higher polaritons are not present because the excitonic pumping is not enough to rise in the states ladder.

\section{Light-matter entanglement}
\label{sec:entanglement}

\begin{figure}[H]
\centering
 \includegraphics[height=4cm, width=7.5cm]{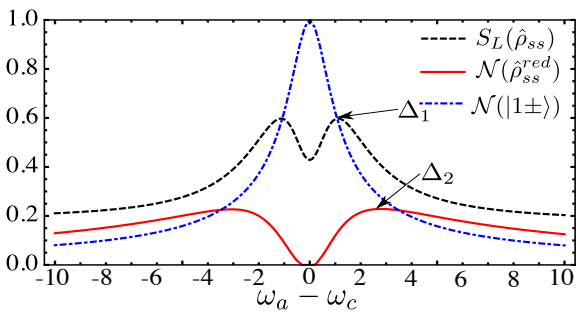}
	\caption{Linear entropy and negativity in the steady state. Parameters: $g_{cm}=g_{am}=20$, $\omega_{c}=1000$, $\omega_{m}=1$, $\kappa=10^{-2}$ and $P_{X}=10^{-3}$. Parameters indicated as $\Delta_{1}$ and $\Delta_{2}$ correspond to the maximum values of linear entropy and negativity, respectively.}
	\label{entanglement}
\end{figure}

Even though vacuum Rabi splitting is a first signature of quantum polaritons, it is necessary to find other physical properties such as light-matter entanglement, non-classical photon statistics, etc. in order to herald genuine quantum polaritons. In this section we focus on a low excitation regime, $\kappa>P_{X}$, to explore single-polariton properties when it is induced by phonons. First of all, we quantify the entanglement between cavity and emitter by computing negativity, which is defined as $\mathcal{N}=2\sum_{\lambda<0}|\lambda|$, where $\lambda$ denotes all the eigenvalues of the partial transpose of the density matrix $\hat{\rho}$ \cite{Horodecki2009}. However, before doing this, a partial trace is carried out on the phonon mode, $\hat{\rho}^{red}=\sum_{\ell}\bra{\ell}\hat{\rho}\ket{\ell}$. The results are shown in Figure \ref{entanglement}. Additionally, the linear entropy, given by $S_{L}=1-Tr(\hat{\rho}^{2})$, is computed in order to compare the purity of the state with the entanglement.

\begin{figure}[H]
\centering
\includegraphics[scale=0.35]{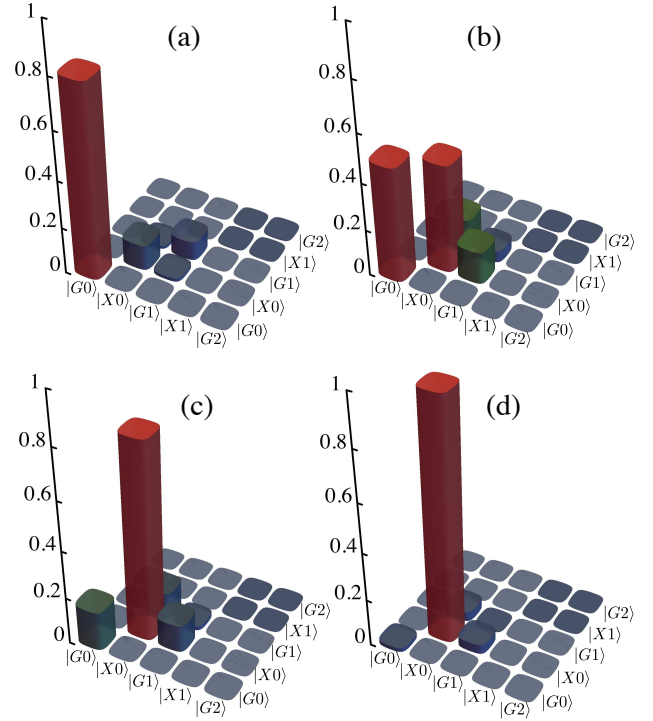}
\caption{Tomography of the reduced density matrix, $|\rho_{ss}^{red}|$ in the steady state for the parameters indicated in Figure \ref{entanglement}. (a) $\Delta=0$, (b) $\Delta=\Delta_1=1.1$, (c) $\Delta=\Delta_2=2.75$ and (d) $\Delta=10$. Other parameters: $g_{cm}=g_{am}=20$, $\omega_{c}=1000$, $\omega_{m}=1$, $\kappa=10^{-2}$ and $P_{X}=10^{-3}$.}
\label{tomography}
\end{figure}

As shown in Figure \ref{entanglement}, when dissipation affects the system, entanglement is not maximum at resonance in comparison with the Hamiltonian case where a polariton state, which in our case is induced by phonons, becomes a Bell state and therefore, maximally entangled. However, Because of the competition between the mechanical interaction and dissipative effects, entanglement is found to be maximum out of resonance. This phenomenology appears also in the Jaynes-Cummings model \cite{Suarez-Forero2016}.  Figure \ref{entanglement} also shows that linear entropy has a local minimum when entanglement vanishes at resonance. To better understand this, in Figure \ref{tomography} it is plotted the density matrix in some key points indicated in Figure \ref{entanglement}: at resonance; $\Delta=0$ (Figure \ref{tomography}a), at maximum of the linear entropy; $\Delta=1.1$ (Figure \ref{tomography}b), at maximum of the negativity; $\Delta=2.75$ (Figure \ref{tomography}c) and for large enough detuning; $\Delta=10$ (Figure \ref{tomography}d).

\vspace{-0.45cm}

At resonance, because losses are larger than gain, most of the population is carried into the ground state with negligible coherences (non-diagonal elements), i.e., a mostly pure and non-entangled state is obtained, $\ket{G,0}$. When linear entropy becomes maximum, a highly mixed state with large entanglement is found. However, in this regime of parameters ($\kappa>P_{X}$), the best condition to finding simultaneously high purity and high entanglement is in $\Delta=2.75$. Finally, very out of resonance, for both Hamiltonian and dissipative cases, entanglement and linear entropy decrease since the mechanical interaction effect is reduced and the combined effect of pumping and losses bring the system to a pure and non-entangled state $\ket{X,0}$.

Furthermore, we find a high fidelity between the phonon-traced density matrix and the Jaynes-Cummings matrix density with the same dissipative rates ($\kappa$ and $P_{X}$, with $P_{X}/\kappa=10^{-1}$) and Rabi coupling strength $g_{ac}=\bar{g}_{m}/2$, $F=Tr\left[\sqrt{\sqrt{\rho_{ss}^{JC}}\rho_{ss}^{red}\sqrt{\rho_{ss}^{JC}}}\right]\approx 1$. This evidences that polariton induced by phonons is very similar to the conventional polariton arising from JCM.

\section{Overview and conclusions}
\label{sec:conclusions}
In this work, we have considered a single mechanical mode interacting linearly with a single mode photonic nanocavity and a two-level single-quantum dot. The phonon mode mediates the coupling between photons in the cavity and excitons in the quantum emitter. By exact diagonalization of the Hamiltonian and then taking into account dissipative effects, we find vacuum Rabi splitting arising from mechanical interactions. This means that our model allows achieving indirect light-matter strong coupling regime just for the phonon mediation. Moreover, it was studied the properties of the phonon-induced polariton, which exhibits the same physics of the Jaynes-Cummings model as is demonstrated by derivation of an effective light-matter Hamiltonian valid for large energy detuning between the mechanical resonator and the cavity (or the emitter). Finally, we studied the light-matter entanglement and purity of the state in a low excitation regime ($P_{X}/\kappa=10^{-1}$) in order to find that the condition with strongest quantum properties; large coherences and low mixing degree, is out of resonance ($\Delta= 2.75$). At resonance, dissipation overcomes the mechanical effect such that the system goes to the ground state and hence, entanglement vanishes. Moreover, it was found almost perfect fidelity between the single-polariton induced by phonons and standard JC polaritons under the same conditions.

The feasibility of this hybrid platform depends mainly on high mechanical interactions. It could be achieved by a large coherent driven which makes part of the underlying theory to our model. Good candidates to implement the model proposed are the quantum dot-microcavity systems taking advantage of the mechanical modes of semiconductor nanostructures and the advanced field of surface acoustic waves in solid state systems. Furthermore, in Circuit Quantum Electrodynamics, superconducting qubits provide high interaction strengths with nanomechanical resonators.

\section*{Acknowledgements}
The authors acknowledge partial financial support from COLCIENCIAS under project project “Emisi\'on en sistemas de Qubits Superconductores acoplados a la radiaci\'on. C\'odigo 110171249692, CT 293-2016, HERMES 31361”. J.E.R.M. thanks financial support from the “Beca de Doctorados Nacionales de COLCIENCIAS 727” and J.P.R.C. is grateful to the “Beca de Doctorados Nacionales de COLCIENCIAS 785”.

\nocite{*}
\bibliographystyle{elsarticle-harv}
\bibliography{Ref.bib}

\appendix

\section{Derivation of the linear cavity-phonon and emitter-phonon Hamiltonian}
\label{app:linearization}

The "linearization" of both interaction Hamiltonians (cavity-phonon and emitter-phonon) is a combination of derivations carried out in \cite{Aspelmeyer2014} and in \cite{Cotrufo2017}. In standard optomechanics, the quantized resonator displacement \cite{Schwab2005}, $\hat{x}$, modifies the cavity frequency because it changes the length of the cavity. In first approximation,  

\begin{equation}
\omega_c(x)\approx\omega_c(0)+\frac{\partial\omega(x)}{\partial x}\bigg\rvert_{x=0}\hat{x}.
\end{equation}

\noindent The mechanical resonator is modeled by a quantum harmonic oscillator, where the displacement operator, $\hat{x}=x_{ZPF}(\hat{b}^{\dagger}+\hat{b})$, is written in terms of annihilation and creation phonon operator. $x_{ZPF}$ is the zero-point fluctuation amplitude of the mechanical resonator. The optomechanical interaction Hamiltonian is then,

\begin{equation}
\hat{H}_{OM}=\hbar g_{OM}\hat{a}^{\dagger}\hat{a}\left( \hat{b}+\hat{b}^{\dagger}\right),
\end{equation}

\noindent where $g_{OM}=(\partial \omega /\partial x) x_{ZPF}$. For more detail see Section III in \cite{Aspelmeyer2014}. Now it is considered a laser driving to the cavity which allows separate the cavity field in two compounds, one average coherent amplitude $\bar{\alpha}$ because of the laser driving, and other fluctuating term $\delta\hat{a}$, which is the one we will be interested. By substituting $\hat{a}=\bar{\alpha}+\delta \hat{a}$ in the interaction Hamiltonian, we have

\begin{equation}
\hat{H}_{OM}=\hbar g_{OM} \left[|\bar{\alpha}|^{2}+\bar{\alpha}\delta\hat{a}^{\dagger}+\bar{\alpha}^{*}\delta\hat{a}+\delta \hat{a}^{\dagger}\delta\hat{a} \right]\left( \hat{b}+\hat{b}^{\dagger}\right).
\end{equation}

\noindent  quadratic term in $\delta\hat{a}$ is negligible respect to the other terms accompanied by $\bar{\alpha}$. The quadratic term in $\bar{\alpha}$ is an average radiation pressure that may be omitted by a shift of the displacement's origin. So, the effective optomechanical Hamiltonian is:

\begin{equation}\label{optomechanical}
\hat{H}_{OM}=\hbar g_{cm} \left(\delta\hat{a}+\delta\hat{a}^{\dagger} \right)\left(\hat{b}+\hat{b}^{\dagger} \right),
\end{equation}

\noindent where $g_{cm}=g_{OM}\sqrt{\bar{n}_{cav}}$ is the optomechanical coupling strength. The condition $g_{cm}>\kappa$ leads to the strong coupling regime in cavity optomechanics \cite{Aspelmeyer2014,Groblacher2009,Verhagen2012}.\\

With regard to the emitter-phonon interaction, the "linearization" of the Hamiltonian, as is proposed in \cite{Cotrufo2017}, starts assuming that mechanical resonator does not affect just the frequency cavity but also the emitter-cavity coupling rate,

\begin{equation}
g(x)\approx g(0)+\frac{\partial g(x)}{\partial x}\bigg\rvert_{x=0}\hat{x},
\end{equation}

\noindent such that the light-matter interaction Hamiltonian, using the rotating wave  approximation (RWA) and assuming the direct light-matter interaction off ($g(0)$=0) yields,

\begin{equation}
\hat{H}_{int}=\gamma \left(\hat{b}+\hat{b}^{\dagger} \right)\left(\hat{\sigma}^{\dagger}\hat{a}+\hat{\sigma}\hat{a}^{\dagger} \right),
\end{equation}

\noindent where $\gamma=(\partial g/\partial x) x_{ZPF}$. Again, considering a laser driving to the cavity, we may express the cavity field operator as $\hat{a}=\bar{\alpha}+\delta\hat{a}$ and now the fluctuation terms can be neglected such that this interaction Hamiltonian becomes,

\begin{equation}\label{atom-phonon}
\hat{H}_{int}=\hbar g_{am} \left(\hat{\sigma}+\hat{\sigma}^{\dagger} \right)\left(\hat{b}+\hat{b}^{\dagger} \right),
\end{equation}

\noindent where $g_{am}=\gamma \sqrt{\bar{n}_{cav}}$ is the atom-phonon coupling rate. This mechanics is referred in the literature as an \textit{atom-phonon interaction through mode field coupling} \cite{Cotrufo2017}. In Hamiltonians (\ref{optomechanical}) and (\ref{atom-phonon}) the coupling rates can be increased by a driving laser to the cavity since they depend on $\bar{\alpha}=\sqrt{\bar{n}_{cav}}$. Also it can be analyzed the role of the rotating wave approximation in these Hamiltonians.

\section{Effective Hamiltonian in a cavity-phonon large detuning}
\label{app:dispersive}

Following an analogous derivation from \cite{Gerry2005c}, a phonon dispersive Hamiltonian is obtained valid in a large cavity-phonon energy detuning regime. Starting from the linear Hamiltonian \ref{total} and with a system of unit where $\hbar=1$, we transform the Schr\"odinger equation to the interaction picture (IP):

\begin{equation}
i\frac{d}{dt}\ket{\Psi_{IP}(t)}=\hat{H}_{IP}\ket{\Psi_{IP}(t)}
\label{schrodinger},
\end{equation}

\noindent by using the unitary operator $\hat{U}=\exp(-i\hat{H}_{0}t)$ over the interaction Hamiltonian: 

\begin{eqnarray}
\hat{H}_{IP}=\hat{U}^{-1} \hat{H}_{int}\hat{U}=g_{cm}\left(\hat{a}^{\dagger}\hat{b} e^{i\Delta_{c}t}+\hat{a}\hat{b}^{\dagger} e^{-i\Delta_{c}t} \right) \nonumber \\
 +g_{am}\left(\hat{\sigma}^{\dagger}\hat{b} e^{i\Delta_{a}t}+\hat{\sigma}\hat{b}^{\dagger} e^{-i\Delta_{a}t} \right),
\label{hamiltonianIP}
\end{eqnarray}

\noindent where $\Delta_{c}=\omega_{c}-\omega_{m}$ and $\Delta_{a}=\omega_{a}-\omega_{m}$ are the cavity-phonon and emitter-phonon energy detuning, respectively. The formal solution of equation (\ref{schrodinger}) is:

\begin{eqnarray}\label{formal}
&&\ket{\Psi_{IP}(t)}=\hat{\mathcal{T}}\left[\exp\left(-i\int_{0}^{t}dt'\hat{H}_{IP}(t')\right) \right]\ket{\Psi_{IP}(0)} \nonumber \\
&&=\left(\hat{1}+(-i)A+\frac{(-i)^{2}}{2}B+\cdots\right)\ket{\Psi_{IP}(0)}.
\end{eqnarray}

\noindent In the last part of this equation was made a perturvative expansion and then was taken into account time-ordering integration. The terms in (\ref{formal}) are:

\begin{equation}
A=\int_{0}^{t}dt'\hat{H}_{IP}(t')
\end{equation}

\noindent and 

\begin{equation}
B=2\int_{0}^{t}dt'\hat{H}_{IP}(t')\int_{0}^{t'}dt''\hat{H}_{IP}(t'').
\end{equation}

\noindent Solving these integrals for $H_{IP}$ in equation (\ref{hamiltonianIP}), we have:

\begin{eqnarray}
\hspace{-0.5cm} A=-i\frac{g_{cm}}{\Delta_{c}}\left(\hat{a}^{\dagger}\hat{b}\left(e^{i\Delta_{c}t}-1\right)-\hat{a}\hat{b}^{\dagger} \left(e^{-i\Delta_{c}t}-1\right) \right) \nonumber \\
\hspace{-0.5cm} -i\frac{g_{am}}{\Delta_{a}}\left(\hat{\sigma}^{\dagger}\hat{b}\left(e^{i\Delta_{a}t}-1\right)-\hat{\sigma}\hat{b}^{\dagger} \left(e^{-i\Delta_{a}t}-1\right) \right).
\end{eqnarray}

\noindent This term can be neglected if the mean excitations, $\braket{\hat{a}^\dagger\hat{b}}\approx\sqrt{\braket{\hat{n}_{a}\hat{n}_{b}}}$ and $\braket{\hat{\sigma}^\dagger\hat{b}}\approx\sqrt{\braket{\hat{n}_{\sigma}\hat{n}_{b}}}$ are not too large, and if $g_{cm}/\Delta_{c}\ll 1$ (and $g_{am}/\Delta_{a}\ll 1$), which is valid for large detuning.\\

Now neglecting all the terms quadratics in $\frac{g_{am/cm}}{\Delta_{a/c}}$, which are small in comparison with terms $\frac{g_{am/cm}^{2}}{\Delta_{c/a}}$ and $\frac{g_{cm}g_{am}}{\Delta_{c/a}}$, because of the large detuning, the integral $B$ yields

\begin{eqnarray}
(i t)^{-1}B=&&\frac{g_{cm}^{2}}{\Delta_{c}}\left( \hat{a}^{\dagger}\hat{a}-\hat{b}^{\dagger}\hat{b}\right)+\frac{g_{am}^{2}}{\Delta_{a}}\left( \hat{\sigma}^{\dagger}\hat{\sigma}+\hat{\sigma}_{z}\hat{b}^{\dagger}\hat{b} \right) \nonumber  \\
&&+ g_{cm}g_{am}\left(\hat{a}^{\dagger}\hat{\sigma}-\hat{a}\hat{\sigma}^{\dagger} \right)\hat{b}\hat{b}^{\dagger}\left(\frac{1}{\Delta_{a}}-\frac{1}{\Delta{c}} \right) \nonumber \\
&&+ g_{cm}g_{am}\left(\frac{1}{\Delta_{a}}\hat{a}\hat{\sigma}^{\dagger}+ \frac{1}{\Delta_{c}}\hat{a}^{\dagger}\hat{\sigma} \right),
\label{B}
\end{eqnarray}

\noindent Rewriting the equation (\ref{formal}) as

\begin{eqnarray}
\ket{\Psi_{IP}(t)}\approx\left(\hat{1}-i t \hat{H}_{eff}\right)\ket{\Psi_{IP}(0)},
\end{eqnarray}

\noindent and comparing with (\ref{B}), we obtain an effective phonon-dispersive Hamiltonian which is Hermitian in cavity-emitter resonance, i.e., $\Delta=\Delta_{c}-\Delta_{a}=0$,

\begin{eqnarray}
\hat{H}_{eff}=\frac{g_{cm}^{2}}{\Delta_{c}}\hat{a}^{\dagger}\hat{a} &+& \frac{g_{am}^{2}}{\Delta_{c}}\hat{\sigma}^{\dagger}\hat{\sigma}+\left(\frac{g_{am}^{2}}{\Delta_{c}}\sigma_z - \frac{g_{cm}^{2}}{\Delta_{c}} \right) \hat{b}^{\dagger}\hat{b} \nonumber\\
& &\frac{g_{cm}g_{am}}{\Delta_c}\left( \hat{a}\hat{\sigma}^{\dagger} + \hat{a}^{\dagger}\hat{\sigma} \right).
\end{eqnarray}

\end{document}